\documentclass[aps,prl,twocolumn,showpacs,superscriptaddress]{revtex4-1}
\usepackage{graphicx}
\usepackage{amsmath}
\usepackage{amssymb}
\usepackage{colordvi}
\usepackage{mathrsfs}
\usepackage{bm}
\usepackage{verbatim}
\usepackage{dcolumn}
\usepackage{bm}
\usepackage{epsfig}
\usepackage{subfigure}
\usepackage{makecell}
\usepackage[colorlinks,allcolors=blue]{hyperref}

\begin{document}
\title{Large Rashba Spin-Orbit Coupling and High-Temperature Quantum Anomalous Hall Effect in Re-Intercalated Graphene/CrI$_3$ Heterostructure}
\author{Yulei Han$^\dag$}
\affiliation{CAS Key Laboratory of Strongly-Coupled Quantum Matter Physics, and Department of Physics, University of Science and Technology of China, Hefei, Anhui 230026, China}
\affiliation{International Center for Quantum Design of Functional Materials, Hefei National Laboratory for Physical Sciences at Microscale, University of Science and Technology of China, Hefei, Anhui 230026, China}
\affiliation{Department of Physics, Fuzhou University, Fuzhou, Fujian 350108, China}
\author{Zhi Yan$^\dag$}
\affiliation{CAS Key Laboratory of Strongly-Coupled Quantum Matter Physics, and Department of Physics, University of Science and Technology of China, Hefei, Anhui 230026, China}
\affiliation{International Center for Quantum Design of Functional Materials, Hefei National Laboratory for Physical Sciences at Microscale, University of Science and Technology of China, Hefei, Anhui 230026, China}
\affiliation{School of Chemistry and Materials Science of Shanxi Normal University $\&$ Key Laboratory of Magnetic Molecules and Magnetic Information Materials of Ministry of Education, Linfen 041004, China}
\author{Zeyu Li}
\affiliation{CAS Key Laboratory of Strongly-Coupled Quantum Matter Physics, and Department of Physics, University of Science and Technology of China, Hefei, Anhui 230026, China}
\author{Xiaohong Xu}
\affiliation{School of Chemistry and Materials Science of Shanxi Normal University $\&$ Key Laboratory of Magnetic Molecules and Magnetic Information Materials of Ministry of Education, Linfen 041004, China}
\author{Zhenyu Zhang}
\affiliation{International Center for Quantum Design of Functional Materials, Hefei National Laboratory for Physical Sciences at Microscale, University of Science and Technology of China, Hefei, Anhui 230026, China}
\author{Qian Niu}
\affiliation{CAS Key Laboratory of Strongly-Coupled Quantum Matter Physics, and Department of Physics, University of Science and Technology of China, Hefei, Anhui 230026, China}
\author{Zhenhua Qiao}
\email[Correspondence author:~~]{qiao@ustc.edu.cn}
\affiliation{CAS Key Laboratory of Strongly-Coupled Quantum Matter Physics, and Department of Physics, University of Science and Technology of China, Hefei, Anhui 230026, China}
\affiliation{International Center for Quantum Design of Functional Materials, Hefei National Laboratory for Physical Sciences at Microscale, University of Science and Technology of China, Hefei, Anhui 230026, China}

\date{\today{}}

\begin{abstract}
  In 2010, quantum anomalous Hall effect (QAHE) in graphene was proposed in the presence of Rashba spin-orbit coupling and ferromagnetic exchange field. After a decade's experimental exploration, the anomalous Hall conductance can only reach about 0.25 in the units of $2e^2/h$, which was attributed to the tiny Rashba spin-orbit coupling. Here, we theoretically show that Re-intercalation in graphene/CrI$_3$ heterostructure can not only induce sizeable Rashba spin-orbit coupling ($>$ 40~meV), but also open up large band gaps at valleys $K$ (22.2 meV) and $K' $ (30.3 meV), and a global band gap over 5.5 meV (19.5 meV with random Re distribution) hosting QAHE. A low-energy continuum model is constructed to explain the underlying physical mechanism. We find that Rashba spin-orbit coupling is robust against external stress whereas a tensile strain can increase the global bulk gap. Furthermore, we also show that Re-intercalated graphene with hexagonal boron-nitride can also realize QAHE with bulk gap over 40~meV, indicating the tunability of $5d$-intercalated graphene-based heterostructure. Our finding makes a great leap towards the experimental realization of graphene-based QAHE, and will definitely accelerate the practical application of graphene-based low-power electronics.
\end{abstract}

\maketitle
\textit{Introduction---.} Quantum anomalous Hall effect (QAHE) is the quantized version of anomalous Hall effect without requiring an external magnetic field~\cite{Haldane,review1,review2}, and is significant in both fundamental physics and potential applications for low-power electronics. In the last decade, QAHE was mainly proposed in materials with Dirac-like dispersion, e.g., $\mathbb{Z}_2$ topological insulator thin films and graphene systems. It is known that the interplay between spin-orbit coupling (SOC) and ferromagnetism is crucial for the QAHE realization~\cite{TI_QAHE1, Gr_QAHE1}. On one side, topological insulators naturally possess considerable SOC, thus one only needs to introduce the ferromagnetism to realize QAHE. Three years after the first theoretical proposal in 2010, $\mathbb{Z}_2$ topological insulator-based QAHE was experimentally observed in magnetically-doped topological insulators at the temperature of 30 mK~\cite{Chang_2013}, and recently the QAHE observation temperature has been raised up to 6.5 Kelvin~\cite{Deng_2020}.

On the other side, although the graphene-based QAHE was also proposed in 2010 by introducing Rashba SOC and ferromagnetism~\cite{Gr_QAHE1}, so far the experimentally observed anomalous Hall conductance can only reach up to 25\% of $2 e^2/h$ by coupling graphene with Y$_3$Fe$_5$O$_{12}$ magnetic thin film~\cite{Gr_Exp1,Gr_Exp2}. It was shown that atomic doping is most effective in inducing Rashba SOC and ferromagnetism due to the chemical bonding between \textit{3d}-orbitals of adatoms and $\pi$-orbital of graphene~\cite{Gr_QAHE1,Gr_adsorption1,Gr_adsorption2}. However, the magnetic adatoms tend to form clusters due to the atomic migration. Later, we found that the magnetic insulating substrate can form a stable heterostructure with graphene~\cite{Gr_BFO}, which can open up a band gap to host QAHE~\cite{Gr_CGT,Gr_CrI3,Gr_NiI2,Gr_AFM}. Its major obstacle is the tiny band gap ($\sim$1 meV) from the weak Rashba SOC that is determined by the physical van der Waals interaction between magnetic insulator and graphene~\cite{Gr_Exp1, Gr_Exp2, Gr_Exp3, Gr_Deng2017}. Thus, pursuing large Rashba SOC becomes crucial for the observation of QAHE in graphene.

By integrating the above two strategies, element intercalation seems to be an ideal choice. So far, there have been various intercalated graphene systems that were proposed in exploring Kane-Mele type $\mathbb{Z}_2$ topological insulator or large Rashba effect in metallic substrate~\cite{Gr_IC_Exp1, Gr_IC_Exp2, Gr_IC1,Gr_IC2,Gr_IC3,Gr_IC_Exp4,Gr_Au_Ni1,Gr_Au_Ni2,Gr_IC_Exp3,BLG_Li,BLG-3d}. Inspired by these findings and recently observed two-dimensional ferromagnetic materials (e.g., CrI$_3$)~\cite{CGT_exp,CrI3_exp,mag_review1,mag_review2}, in this Letter, we provide a systematic study on $5d$-intercalated graphene/CrI$_3$ systems. We find that the large Rashba SOC over 40 meV can be induced, and a sizeable topological band gap opens up to host the QAHE with a Chern number of $\mathcal{C}=-2$. We implement an effective model to clearly understand the effects from exchange field, intrinsic and Rashba SOC in the formation of QAHE. By applying external stress, we further show that the Rashba gap is robust whereas a tensile strain can even enhance the bulk band gap. Furthermore, we also find that large Rashba gap, as well as QAHE bulk gap over 40 meV, can be realized by intercalating Re atoms between graphene and hexagonal boron nitride, implying the tunability of $5d$-intercalated graphene-based system. Our findings provide a solid strategy to engineer large Rashba SOC and therefore open a sizable band gap to harbor QAHE, as well as provide a concrete recipe for the experimental realization of QAHE in graphene.

\begin{figure}
  \centering
  \includegraphics[width=8.0cm]{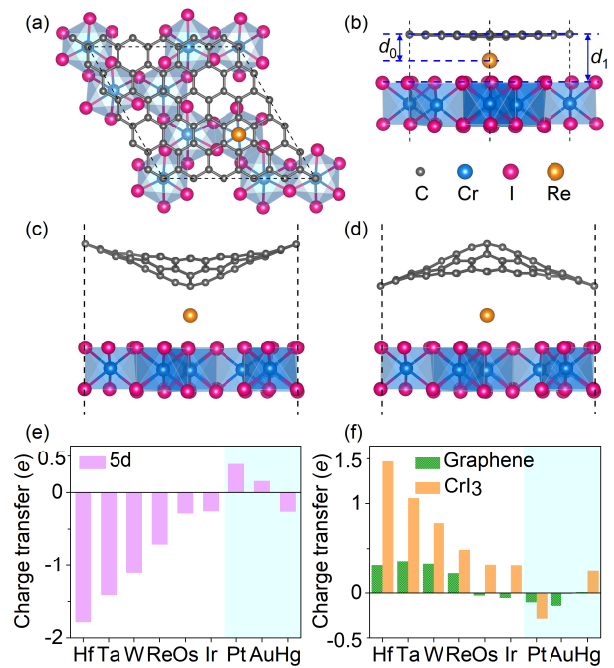}
  \caption{(a) Top and (b) side views of the Re-intercalated graphene/CrI$_3$ heterostructure. $d_0$ represents the average distance between Re atom and the surrounded six carbon atoms, and $d_1$ represents the average distance between graphene and CrI$_3$. (c)-(d) Schematic plot of  $5d$-intercalation induced (c) attractive and (d) repulsive interaction between $5d$ atom and graphene. (e)-(f) Charge transfer for (e) $5d$ atom and (f) graphene/CrI$_3$ in the intercalated systems.}\label{Fig1}
\end{figure}

\textit{System Models and Calculation Methods---.}
Our first-principles calculations are performed by using the projected augmented-wave method~\cite{PAW} as implemented in the Vienna $ab~initio$ simulation package (VASP)~\cite{VASP1,VASP2} and the detailed parameters are described in Supplemental Materials~\cite{SM}.
We first investigate the stable intercalation site of $5d$ atoms and take Re as an example. Four different heterostructure configurations, i.e., C(H)-Re-Cr(T), C(T)-Re-Cr(T), C(T)-Re, and C(T)-Re-I(H) are constructed, with H and T respectively representing ``hollow" and ``top" sites [See Figs.~\ref{Fig1}(a) and \ref{Fig1}(b) for the first configuration and Fig.~S1 for other three configurations]~\cite{SM}.
We find that C(H)-Re-Cr(T) is the most stable geometry for Re intercalation from the total energy calculations as summarized in Table S1~\cite{SM}. Table~\ref{Tab2} summarizes the atomic distances $d_0$ and $d_1$ of $5d$-intercalated graphene/CrI$_3$ heterostructure as defined in Fig.~\ref{Fig1}(b). One can see that $d_0$ and $d_1$ for Re-intercalated system are respectively 1.78 and 3.59 \AA, and also are the shortest among all $5d$-intercalated system, indicating the strongest Re-mediated van der Waals interaction between graphene and CrI$_3$. The inclusion of intercalated atoms may induce the attractive or repulsive interaction as schematically displayed in Figs.~\ref{Fig1}(c) and \ref{Fig1}(d). By analyzing the relative positions of the surrounded carbon atoms, we find an \emph{attractive} interaction between $5d$ atom and graphene for Hf-Ir, whereas a \emph{repulsive} interaction for Pt/Au/Hg [see Table~\ref{Tab2} and Fig.~S2]. This can be qualitatively understood by the charge transfer process as demonstrated in Figs.~\ref{Fig1}(e) and \ref{Fig1}(f), with a sign change from Ir to Pt for $5d$ atoms and CrI$_3$.
Hereinbelow, we focus on exploring the electronic properties of a concrete Re-intercalated graphene/CrI$_3$ heterostructure.

\begin{figure}
  \centering
  \includegraphics[width=8cm]{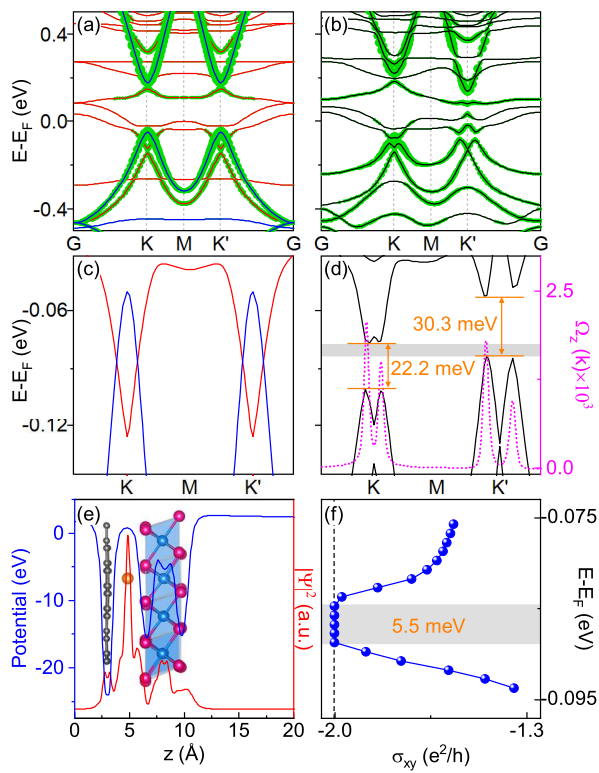}
  \caption{(a)-(b) Electronic band structures of Re-intercalated graphene/CrI$_3$ heterostructure (a) without SOC and (b) with SOC. Red and blue in (a) represent spin-up and spin-down states, respectively. Green circle represents states from C-$p_z$ orbital. (c)-(d) The corresponding zooming-in band structures near the band crossing point. In (d), the Berry curvature and the global band gap are shown by pink dashed line and gray rectangle, respectively. (e) The planar-averaged distribution of potential and electron density $|\Psi (z)|^2$ along out-of-plane direction. The corresponding atomic positions are also superposed onto the curves. (f) Anomalous Hall conductance as a function of Fermi level, with $\sigma_{xy}$ being quantized for energies inside the bulk gap.}\label{Fig2}
\end{figure}

\textit{Band Structures---.} We start from analyzing the band structures. Figures~\ref{Fig2}(a) and \ref{Fig2}(c) display the spin-resolved bands, with C-$p_z$ orbital contribution being highlighted in green. One can find large spin splitting of $\sim$300 meV/94 meV in conduction/valence bands, and can also observe that the Re-intercalation greatly enhances the exchange interaction between CrI$_3$ and graphene from spin density distribution [see Figs. S3(a) and S3(b)]~\cite{SM}. The crossing of spin-up and spin-down bands indicates a possible band gap opening if Rashba SOC can be introduced as proposed~\cite{Gr_QAHE1}. The Heisenberg exchange constant and Curie temperature from mean-field theory~\cite{CrI3_Curie} are estimated to be 11.73 meV and 204 K, respectively. The remarkable increase of Curie temperature from Re intercalation in graphene/CrI$_3$ heterostructure makes it a feasible platform for realizing high-temperature QAHE.

As displayed in Figs.~\ref{Fig2}(b) and \ref{Fig2}(d), the inclusion of spin-orbit coupling leads to a sizeable local band gap of 22.2/30.3 meV at valley $K$/$K'$. It is noteworthy that the local band gaps at valleys K and K' are not aligned, indicating the potential presence of other physical ingredients, e.g., the sublattice staggered potential or Kane-Mele type intrinsic SOC. Even so, a considerable global band gap around 5.5 meV still exists, which is about five times larger than our previous theoretical predictions in graphene/BiFeO$_3$ systems~\cite{Gr_BFO}.
By analyzing different atomic orbitals [see Fig. S4]~\cite{SM}, one can see that Cr-($d_{xy},~d_{x^2-y^2},~d_{yz},~d_{xz}$), I-$p$, C-$p_z$ and Re-$d_{z^2}$ orbitals are strongly hybridized near the global gap. The inclusion of Re atom hybridizes C-$p_z$ and I-$p$ orbitals via Re-$d_{z^2}$ orbital, leading to the enhancement of exchange interaction and Rashba SOC. This can also be clearly verified from the charge transfer, i.e., Re atom transfers 0.22/0.49 electron to C/I, and the presence of Re intercalation changes the charge distribution dramatically and enhances the charge density overlap between CrI$_3$ and graphene [see Figs. S3(c)-S3(d)]. %The slight electron doping in graphene can also be observed in Figs.~\ref{Fig2}(c) and \ref{Fig2}(d), where the band crossing point is around -0.08 eV.
Moreover, the enhancement of Rashba SOC can be qualitatively explained by the planar-averaged asymmetric potential and electron density distributions [see Fig.~\ref{Fig2}(e)],  with the screening charge around the band crossing point primarily distributed around Re, implying the critical role of Re-intercalation in increasing Rashba SOC.

\begin{table}
 \caption{Atomic distance of $5d$-intercalated graphene/CrI$_3$ heterostructure. The most stable geometry of C(H)-X-Cr(T) is used. The last row denotes the attractive (+) / repulsive (-) interaction between graphene and $5d$ atoms.
 }\label{Tab2}
 \begin{ruledtabular}
  \begin{tabular}{ccccccccccc}
   X = & Hf & Ta & W & Re & Os & Ir & Pt & Au & Hg \\ \hline
   $d_0$  & 2.15 & 1.98 & 1.88 & 1.78 & 2.22 & 1.89 & 3.61 & 3.52 & 3.86  \\
   $d_1$  & 3.66 & 3.61 & 3.59 & 3.59 & 3.69 & 3.65 & 3.70 & 3.70 & 3.69  \\
   interaction & + & + & + & + & + & + & - & - & - \\
  \end{tabular}
 \end{ruledtabular}
\end{table}

\textit{Topological Properties---.} Then, we study the topological properties of the Re-intercalated graphene/CrI$_3$ system. As shown in Fig.~\ref{Fig2}(d), one can observe that the Berry curvature exhibits large peaks around valleys $K$ and $K'$, but vanishes elsewhere, indicating the existence of nonzero Chern number $\mathcal{C}$, i.e., the formation of QAHE. To quantitatively verify this finding, we calculate the anomalous Hall conductance $\sigma_{xy}$. Figure~\ref{Fig2}(f) displays that the Hall conductance is quantized when the Fermi energy lies inside the band gap, i.e., $\sigma_{xy}=-2.0 ~e^2/h$, which is a strong evidence of QAHE.

\begin{figure}
  \centering
  \includegraphics[width=8.0cm]{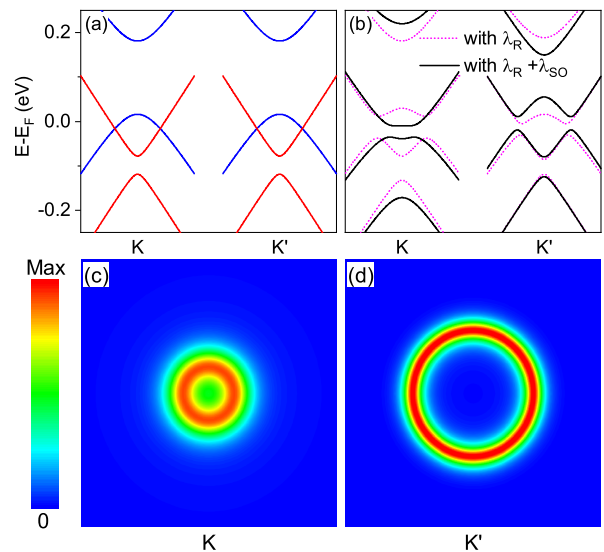}
  \caption{(a)-(b) Band structures from effective model (a) without SOC and (b) with SOC, respectively. The red (blue) color in (a) denotes spin-up (spin-down) state. The pink dotted line in (b) represents that only Rashba SOC is included whereas the black line represents that both Rashba and Kane-Mele SOCs are considered. (c)-(d) The corresponding Berry curvature distributions around (c) $K$ and (d) $K'$ points with SOC.}\label{Fig3}
\end{figure}

The Re-intercalated graphene/CrI$_3$ system is distinct from the previously studied graphene/CrI$_3$ and $5d$-adsorbed graphene.
In the previous graphene/CrI$_3$ heterostructure~\cite{Gr_CrI3}, the formation of QAHE with a small band gap requires a compressive strain to shift the Dirac cone of graphene to the gap of CrI$_3$. In contrast, the realization of QAHE in our Re-intercalated system does not require external stress and intrinsically has large Rashba SOC. %indicating that graphene-based QAHE is feasible to be experimentally implemented.
In $5d$-adsorbed graphene, the strong hybridization between $d_{xz/yz}$ orbital and C-$p_z$ orbital plays a crucial role in the formation of various topological phases~\cite{Hu_2012, Zhang_2012}.
To demonstrate the differences between Re-intercalated and Re-adsorbed systems, we compare the bands and density of states with/without CrI$_3$ in Re-intercalated graphene/CrI$_3$. As displayed in Fig. S5, in the absence of CrI$_3$, strong hybridization between C-$p_z$ and Re-($s$, $d_{xz/yz}$) orbitals is around the Fermi level whereas Re-$d_{z^2}$ orbital is far away from the Fermi level, which is different from the density of states distribution in Re-intercalated graphene/CrI$_3$ [see Fig. S4]~\cite{SM}. Therefore, we can find that the presence of CrI$_3$ significantly modifies the band structure of the heterostructure in two aspects, i.e., (i) the $d_{xz/yz}$ and $d_{z^2}$ orbitals of Re are moved up and (ii) C-$p_z$ orbital is strongly hybridized with $d_{z^2}$ orbital, leading to a distinct physical origin of topological phases.

\textit{Low-Energy Continuum Model---.} The electronic states around valleys $K/K'$  are dominated by $p_z$ orbital of graphene. Therefore, one can construct an effective model Hamiltonian as below~\cite{Gr_QAHE1,Gr_BFO}:
\begin{widetext}
\begin{eqnarray}
 H(\mathbf{k})=v_{\mathrm{F}}(\eta \sigma_{x}k_{x} + \sigma_{y}k_{y})\mathbf{1}_{S}+ \frac{\lambda_{R}}{2}(\eta \sigma_{x}s_{y}-\sigma_{y}s_{x})+ \eta \lambda_{SO}\sigma_z s_z + m_{a}\frac{\mathbf{1}_{\sigma}+\sigma_z}{2} + m_{b}\frac{\mathbf{1}_{\sigma}-\sigma_z}{2} +\delta \sigma_{z}\mathbf{1}_{S},
\end{eqnarray}\label{eq1}
\end{widetext}
where $v_{\mathrm{F}}$ is the Fermi velocity, $\eta = \pm 1$ labels the valley index, $\mathbf{\sigma}$ and $\mathbf{s}$ are Pauli matrices for sublattice and spin degrees of freedom, respectively. The first term represents electron's hopping of pristine graphene, the second and third terms are respectively Rashba and intrinsic SOCs. The fourth and fifth terms describe the sublattice-dependent magnetic exchange fields, and the last term is the staggered sublattice potential. By fitting the first-principles band structure without and with SOC, we can extract the effective system parameters, i.e., $m_a=150~\mathrm{meV}$, $m_b = 47~\mathrm{meV}$,  $\delta = 31~\mathrm{meV}$, $\lambda_{SO} = 39 ~\mathrm{meV}$ and $\lambda_{R} = 45~\mathrm{meV}$.

Let us now analyze the influence of each physical ingredient on the band structure. As displayed in Fig.~\ref{Fig3}(a), the presence of unequal exchange field in $A/B$-sublattice leads to a different spin splitting of conduction/valence bands, whereas the two valleys are identical without SOC. It is noteworthy that the unequal exchange field is consistent with the spin density as displayed in Figs. S3(a) and S3(b), where visible spin polarization only exists at one graphene sublattice. When Rashba SOC is further considered [see Fig.~\ref{Fig3}(b), pink dotted line], one can observe that band gaps are opened at the band crossing points at each valley, and the local gaps at $K/K'$ valleys become different due to the presence of staggered sublattice potential. In this case, the global band gap is only determined by the local gap of valley $K'$. However, from Fig.~\ref{Fig2}(d), we can observe that global band gap is determined by conduction band of valley $K$ and valence band of valley $K'$. To well capture this character, the intrinsic SOC $\lambda_{SO}$ should be further included in the effective model. As displayed in Fig.~\ref{Fig3}(b) [black line], the band structure with the four essential physical ingredients reproduces the low-energy physics of the first-principles results.

Figures.~\ref{Fig3}(c) and \ref{Fig3}(d) display the Berry curvature distributions around valley $K$ and $K'$, respectively. The sum over of Berry curvature in each valley gives $\mathcal{C}_{K/K'} = 1$, resulting in Chern number of $\mathcal{C}=\mathcal{C}_{K}+\mathcal{C}_{K'}=2$. It is noteworthy that the Berry curvature distributions are different at two valleys, i.e., the Berry curvature peaks are close to valley $K$ but are far away from valley $K'$, which is also consistent with the Berry curvature from first-principles method [see Fig.~\ref{Fig2}(d)]. Remarkably, the large Rashba SOC strength, i.e., $\lambda_{R}=45~\mathrm{meV}$, in our intercalated system is over 30 times larger than that reported in the previous study~\cite{Gr_BFO}, and the great enhancement of Rashba SOC solves a crucial problem in experimental observation of QAHE in graphene~\cite{Gr_Exp2}. Furthermore, in realistic systems, the effect of staggered sublattice potential vanishes due to the absence of periodicity, leading to the increase of global band gap to be around 19.5 meV [see Fig. S6].

\begin{figure}
  \centering
  \includegraphics[width=8.0cm]{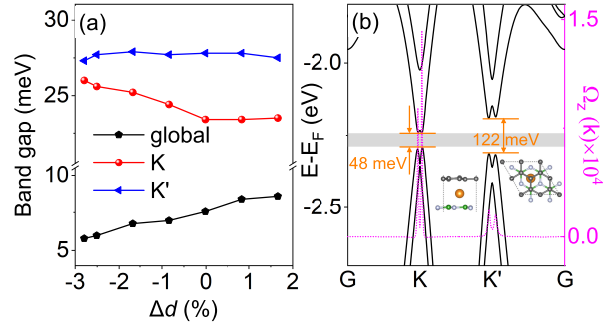}
  \caption{(a) Band gaps of Re-intercalated graphene/CrI$_3$ as a function of out-of-plane strain . (b) Band structure and Berry curvature along high symmetry line in Re-intercalated graphene/\textit{h}-BN system.The inset shows the corresponding configuration.}\label{Fig4}
\end{figure}

\textit{Tunability of 5d-intercalated systems---.}
From the above analysis, we know that although the local gap of each valley is sufficiently large, the global gap is determined by the band gap alignment of the two valleys. One effective approach to tune the band gap is to apply external stress. As displayed in Fig.~\ref{Fig4}(a), the global band gap of Re-intercalated graphene/CrI$_3$ gradually increases (decreases) when applying tensile (compressive) strain, while the local gap of valley $K$/$K'$ still maintains a large value ($>$ 20 meV). Another approach is to change substrate materials. Besides CrI$_3$, other magnetic materials can be utilized as substrates, which requires more exploration.
To our surprise, we find that nonmagnetic materials, e.g., hexagonal boron nitride ($h$-BN), can also induce large spin splitting and Rashba SOC in graphene, and a nontrivial global gap (48 meV) can be observed in Fig.~\ref{Fig4}(b) [see Supplemental Material for more discussion].

\textit{Summary---.}
We have demonstrated that high-temperature QAHE can be realized in $5d$-intercalated graphene-based heterostructures due to the formation of large Rashba SOC. Taking CrI$_3$ as an example, we find that Re-intercalated graphene/CrI$_3$ system harbors large local gaps greater than 20 meV and a global band gap of 5.5 meV exhibiting QAHE with Chern number of $\mathcal{C}=-2$. By constructing an effective model, we show that the Rashba SOC approaches about 45 meV, which is over 30 times larger than that in graphene/BiFeO$_3$~\cite{Gr_BFO}. The relatively small global band gap by the unaligned local gaps of valleys $K/K'$ is attributed to the presence of staggered sublattice potential and intrinsic SOC. In realistic systems with large scale random intercalation, the random distribution of sublattice potentials leads to the formation of an enhanced global gap of 19.5 meV. We further show that applying tensile strain can increase the bulk gap whereas preserve the local gap at valley $K/K'$. Finally, we demonstrate that $5d$ intercalation in graphene with nonmagnetic substrate $h$-BN can induces extremely large global gap ($>40$ meV) to realize QAHE. Our work paves a practical way to realize QAHE in graphene-based system, and may stimulate more theoretical and experimental explorations to search more suitable substrates and intercalated elements in graphene.

\begin{acknowledgments}
This work was financially supported by the National Natural Science Foundation of China (No. 11974327 and No. 12004369), Fundamental Research Funds for the Central Universities (WK3510000010, WK2030020032), Anhui Initiative in Quantum Information Technologies. The Supercomputing services of AM-HPC and USTC are gratefully acknowledged.
\end{acknowledgments}

$^\dag$ These authors equally contribute to this work.

\end{document}